\documentclass[aps,pra,reprint,superscriptaddress]{revtex4-1}

\usepackage{graphicx}
\usepackage{color}
\usepackage[colorlinks=true, pdfstartview=FitV, linkcolor=blue, citecolor=blue, urlcolor=blue]{hyperref}
\usepackage{color}
\usepackage{multirow}
\usepackage{enumerate}
\begin{document}


\title{Quantum anomalous Hall effect in stable 1T-YN$_2$ monolayer with a large nontrivial band gap and high Chern number}


\author{Xiangru Kong}
\affiliation{International Center for Quantum Materials, Peking University, and Collaborative Innovation Center of Quantum Matter, 100871 Beijing, China}

\author{Linyang Li}
\email[]{linyang.li@uantwerpen.be}
\affiliation{Department of Physics, University of Antwerp, Groenenborgerlaan 171, B-2020 Antwerp, Belgium}

\author{Ortwin Leenaerts}
\affiliation{Department of Physics, University of Antwerp, Groenenborgerlaan 171, B-2020 Antwerp, Belgium}

\author{Weiyang Wang}
\affiliation{Department of Physics, University of Antwerp, Groenenborgerlaan 171, B-2020 Antwerp, Belgium}
\affiliation{School of Physics and Electronics Information, Shangrao Normal University, 334001 Shangrao, Jiangxi, China}

\author{Xiong-Jun Liu}
\affiliation{International Center for Quantum Materials, Peking University, and Collaborative Innovation Center of Quantum Matter, 100871 Beijing, China}

\author{Fran\c{c}ois M. Peeters}
\affiliation{Department of Physics, University of Antwerp, Groenenborgerlaan 171, B-2020 Antwerp, Belgium}



\begin{abstract}
The quantum anomalous Hall (QAH) effect is a topologically nontrivial phase, characterized by a non-zero Chern number defined in the bulk and chiral edge states in the boundary. Using first-principles calculations, we demonstrate the presence of the QAH effect in 1T-YN$_2$ monolayer, which was recently predicted to be a Dirac half metal without spin-orbit coupling (SOC).  We show that the inclusion of SOC opens up a large nontrivial band gap of nearly $0.1$ eV in the electronic band structure. This results in the nontrivial bulk topology which is confirmed by the calculation of Berry curvature, anomalous Hall conductance and the presence of chiral edge states. Remarkably, a QAH phase of high Chern number $C = 3$ is found, and there are three corresponding gapless chiral edge states emerging inside the bulk gap.
Different substrates are also chosen to study the possible experimental realization of the 1T-YN$_2$  monolayer while keeping its nontrivial topological properties.
Our results open a new avenue in searching for QAH insulators with high temperature and high Chern numbers, which can have nontrivial practical applications.
\end{abstract}


\maketitle

\section{Introduction}
The discovery of quantum Hall effect brought about a new fundamental concept, the topological phase of matter, to condensed matter physics~\cite{QHE1980}. The quantum Hall effect is obtained in a two-dimensional (2D) electron gas in the presence of a strong perpendicular external magnetic field, which drives the electrons to fill in the discrete Landau levels, resulting in the quantized Hall conductance. Nevertheless, Landau levels are not the necessary ingredient for the realization of the quantum Hall effect, where the integer Hall plateaus are actually interpreted by Chern numbers, a type of topological invariants defined in the 2D momentum space~\cite{thouless1982}. The topological interpretation implies that the realization of quantum Hall effect may be achieved without external magnetic field.
The first toy model for the quantum Hall effect without Landau level, i.e. the quantum anomalous Hall (QAH) effect was proposed by Haldane in 1988 in a 2D honeycomb lattice with next-nearest-neighboring hopping modulated by staggered flux~\cite{Haldane1988PRL}. While there has been theoretical studies of QAH effect in solid state materials~\cite{Onoda2003PRL,Qi2006PRB}, little progress was made in experiment until the discovery of time-reversal (TR) invariant topological insulators~\cite{Hasan2010rmp,Qi2011rmp}. Realization of QAH effect combines several basic ingredients that the system should be in 2D regime, insulating, TR symmetry breaking with a finite magnetic ordering, and has a non-zero Chern number in the valence bands~\cite{Weng2015review,Yang2016cpb}. Many theoretical schemes for realizing QAH insulators have been proposed, including the magnetically doped quantum well-based 2D topological insulators~\cite{LiuCX2008PRL}, graphene with transition metal (TM) adatoms (3$d$, 4$d$ and 5$d$) \cite{Qiao2010PRB,ZhangHB2012PRL,QAHE2014PRB,QAHE2015nanolett}, buckled honeycomb-lattice systems of group IV or V elements with TM adatoms \cite{Ezawa2012PRL,ZhangHB2012PRB,ZhangHB2013PRB}, half-functionalized honeycomb-lattice systems \cite{HS2014PRB,YanBH2014PRL}, heterostructure quantum wells \cite{Vanderbilt2013PRL,nanolett2015well}, organic metal frameworks \cite{LiuF2013PRL}, $sd^2$ `graphene' with TM atoms~\cite{PRL2014sd2} and also ultracold atom systems~\cite{WuC2008PRL,LiuXJ2010PRA}.
In particular, following the theoretical work by Yu \textit{et al.}~\cite{Yu2010QAH}, the QAH effect was first realized experimentally using Cr-doped magnetic thin-film topological insulator, with the quantized Hall conductance being observed~\cite{Chang2013science}. The QAH states have also been reported in experiment for ultracold atoms, with the Haldane model~\cite{Esslinger2014} and a minimized spin-orbit coupled model~\cite{wu2016realization} being realized, respectively. Realization of QAH phases is of great interests both in fundamental theory and potential applications. For example, the chiral edge states of a spin-orbit coupled QAH insulator may exhibit novel topological spin texture in real space, which can have applications in designing spin devices~\cite{LXJ2014PRL}. More interestingly, the heterostructure formed by a QAH insulator and conventional $s$-wave superconductivity can realize chiral topological superconductors, which host Majorana zero modes binding to vortices and chiral Majorana edge modes in the boundary~\cite{Qi2010PRB,LiuXJ2014PRL}. Remarkably, a recent experiment observed such chiral Majorana edge modes based on the QAH insulator/$s$-wave superconductor heterostructure, where a $e^2/(2h)$-plateau of tunneling conductance was observed~\cite{HeQL2017Science}.

Albeit the important experiment progress, there are several limitations for the current study of real QAH materials. First, so far the QAH phases in solid state materials have only be observed at quite low temperature, typically in the order of $mK$, due to small topological bulk gap~\cite{Chang2013science}. To enable broad studies of fundamental physics and potential future applications, it is important to have QAH insulators with a large bulk gap and observable with high temperature.
Furthermore, the current experimental studies are focused on the QAH states with low Chern number $C=\pm1$. The QAH insulators with high Chern numbers exhibit different topological phases which are expected to bring about new fundamental physics and interesting applications, but are yet to be demonstrated experimentally. This motivates us to search for new QAH insulators based on the transition metal compounds.

Transition metal is defined as an element whose atom has a partially filled $d$ sub-shell or which can give rise to cations with an incomplete $d$ sub-shell \cite{1997compendium}.
Due to the partially filled $d$ shell,  TMs can have many different oxidation states when forming compounds and show many appealing  electronic, magnetic, and catalytic properties.
Strong electronegative elements (groups V, VI, and VII) are easily combined with TMs to form stable compounds.
One of the most attractive compounds are the transition metal dichalcogenides (TMDs) with chemical composition TMX$_2$, where TM stands for the TM and X is a chalcogen element such as S, Se, or Te \cite{CSR2015,CSR2015YaoYG}.
TMD monolayers have many new physical properties as compared to their bulk counterparts due to their reduced dimensions. As confirmed by experiment, TMDs can exhibit three different structures, called 1H, 1T, and 1T${'}$ \cite{Qian2014QSHE}.
Using first-principles calculations, TMD monolayers have been predicted to show both semiconducting (1H-MoS$_2$) \cite{MoS2010PRL} and metalic (1T-PtSe$_2$) \cite{Gao2015nano} properties, spin polarization effect (1H-VS$_2$) \cite{Ma2012ACSnano} and quantum spin Hall (QSH) effect (1T${'}$-WTe$_2$) \cite{Qian2014QSHE,zheng2016quantum}, where some of them have been realized experimentally \cite{CSR2015,CSR2015YaoYG}.
However, the QAH effect has not yet been predicted in TMD structures.
A recent expansion of the TMD-like compounds has been realized through MoN$_2$, a nitrogen-rich TM nitrides (TMN) that has been synthesized through a solid-state ion-exchange reaction under high pressure \cite{MON2015JACS}. First-principles calculations predict that 1H-MoN$_2$ monolayer is a high temperature 2D ferromagnetic (FM) material with Curie temperature of nearly 420 K \cite{MON2015Nanolett}. This suggests that combining TM and N atoms can lead to stable 2D monolayers exhibiting novel band structures.
A natural question arises: can we find the QAH effect in the TMNs?

Previous calculations focused on TMN monolayers with the chemical composition TMN$_2$ (TM = Y, Zr, Nb and Tc) which demonstrated that the most energetically and dynamically stable phase is 1T \cite{Wu2016PLA,JMCC2016,Liu2017Nanores}.
Interestingly, the strong nonlocal $p$ orbitals of the N atoms in 1T-YN$_2$ result in a Curie temperature of 332 K.
The three unpaired electrons in the two N atoms give 1T-YN$_2$ a FM ground state with a total magnetic moment of 3 $\mu_B$ per unit cell.
Remarkably, the electronic band structures as obtained from DFT calculations show that 1T-YN$_2$ is a $p$-state Dirac half metal (DHM) in the absence of spin-orbit coupling (SOC).
A DHM is defined as a metal in which a Dirac cone exists at the Fermi level in one spin channel and a band gap opens in the other channel \cite{WangXL2008PRL,Wang2017NSR}.
The 100\% spin-polarization and massless Dirac fermions in DHMs attracted a lot of  attention due to  potential applications in high-speed spintronic devices \cite{Liu2017Nanores}.
In this paper, we investigate the stable 1T-YN$_2$ monolayer using first-principles calculations with the inclusion of SOC. We obtain a relatively large nontrivial band gap ($\approx$ 0.1 eV) in the Dirac cone. The nontrivial properties of the 1T-YN$_2$ monolayers are further confirmed by the calculation of the Berry curvature, the anomalous Hall conductance (AHC), the Chern number, and the corresponding edge states. The large nontrivial band gap and high Chern number are very interesting for practical applications in future nanodevices.

\section{Computational Methods}
The first-principles calculations were done with the Vienna \textit{ab initio} simulation package (VASP) using the projector augmented wave (PAW) method in the framework of Density Functional Theory (DFT) \cite{PhysRevB.54.11169,PhysRevB.48.13115,Kresse1999}. The electron exchange-correlation functional was described by the generalized gradient approximation (GGA) in the form proposed by Perdew, Burke, and Ernzerhof (PBE) \cite{Perdew1996}. The structure relaxation considering both the atomic positions and lattice vectors was performed by the conjugate gradient (CG) scheme until the maximum force on each atom was less than 0.01 eV/\AA, and the total energy was converged to $10^{-5}$ eV with Gaussian smearing method.
To avoid unnecessary interactions between the YN$_2$ monolayer and its periodic images, the vacuum layer is set to at least 17 \AA.
The energy cutoff of the plane waves was chosen as 500 eV. The Brillouin zone (BZ) integration was sampled by using a 31 $\times$ 31$\times$ 1 $\Gamma$-centered Monkhorst-Pack grid. To obtain a more reliable calculation for the electronic band structure, especially the band gap, the screened Heyd-Scuseria-Ernzerhof Hybrid functional method (HSE06) \cite{hse06,hse06e} with mixing constant 1/4 was also used with a 15 $\times$ 15 $\times$ 1 $\Gamma$-centered Monkhorst-Pack grid for BZ integration. 
SOC is included by a second variational procedure on a fully self-consistent basis.
In the calculations of van der Waals (vdW) heterostructures and quantum wells in our work, the zero damping DFT-D3 method of Grimme within the PBE functional was employed for the vdW correction~\cite{jcp2010dftd3}.
An effective tight-binding Hamiltonian constructed from the maximally localized
Wannier functions (MLWF) was used to investigate the surface states \cite{Mostofi20142309,kong2017prb}.
The iterative Green's function method \cite{JPFMP1985} was used with the package WannierTools \cite{wanntools}.

\section{Results}
Spin-orbit coupling (SOC) is a relativistic effect which describes the interaction of the spin of an electron with its orbital motion \cite{dresselhaus2007group,winkler2003spin}.
Though SOC is a small perturbation in a crystalline solid and has little effect on structure and energy, it plays a more important role in the band structure near the Fermi level in the case of  heavy elements.
SOC will cause a spin splitting of the energy bands in inversion-asymmetric systems, and more importantly, it can result in a band opening and a band inversion that gives rise to fascinating phenomena.
Previous studies neglected SOC in 1T-YN$_2$, although the SOC effect of the Y and N atoms are not negligible and turns out to be important to realize the topological properties discussed in our work.
In the following, we will take SOC into account and study the electronic band structure of stable 1T-YN$_2$ monolayer by first-principles calculations.
Though the calculation demonstrated that the easy magnetic axis of 1T-YN$_2$ monolayer is in-plane, the Magnetocrystalline Anisotropy Energy (MAE) of the out-of-plane spin orientation is only about 0.025 meV per spin of every nitrogen atom. Therefore, The spin orientation in the calculations is chosen in the direction of out-of-plane.

\begin{figure}[htb!]
\includegraphics[scale=0.3]{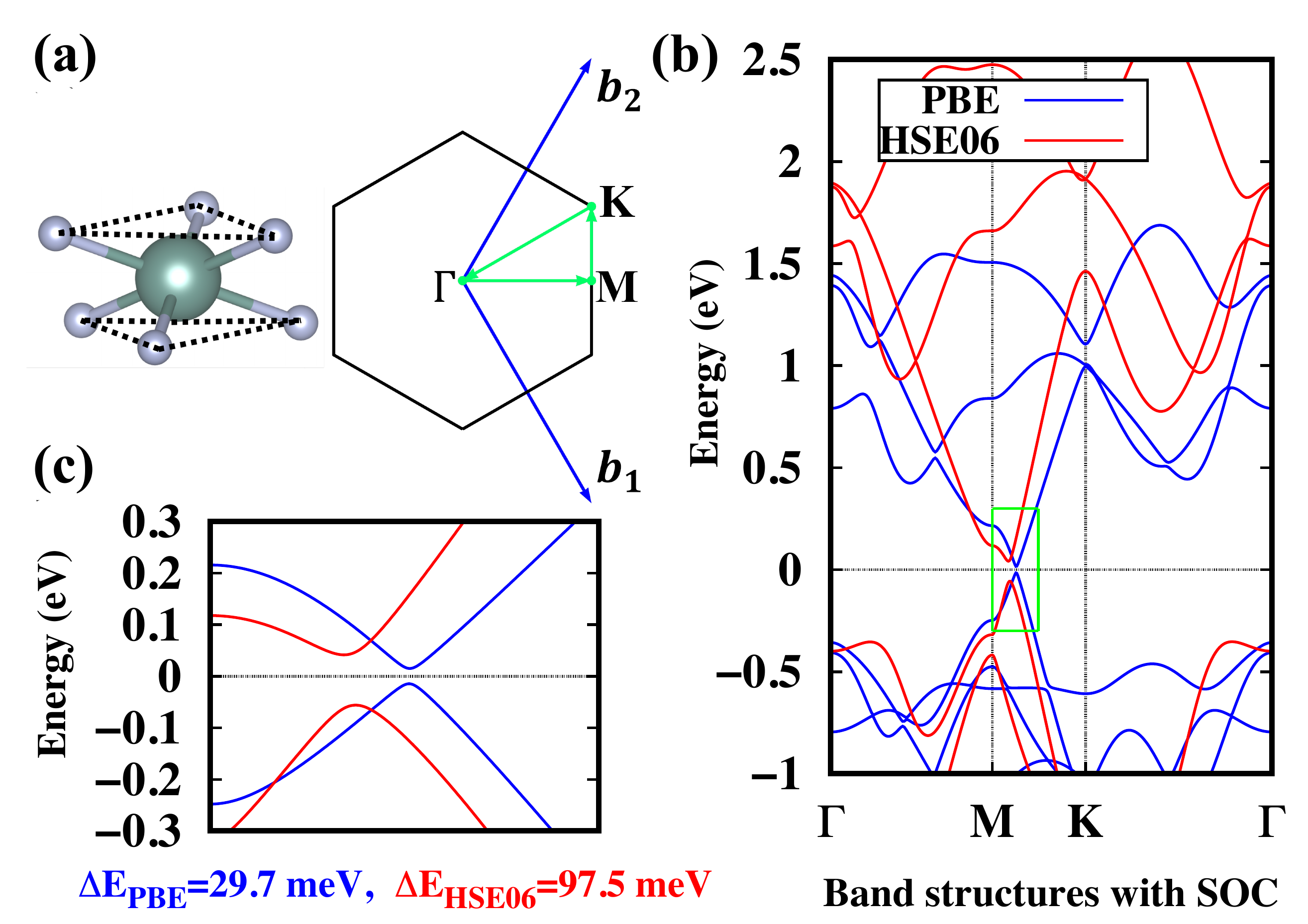}%
\caption{
\label{bz_bands} 	
(a) The octahedral structure unit of 1T-YN$_2$ monolayer and the corresponding 2D Brillouin zone with its high symmetry points ($\Gamma$, M and K). $b_1$ and $b_2$ are the reciprocal lattice vectors. The band structure is calculated along the green path ($\Gamma$-M-K-$\Gamma$).
(b) Band structures of 1T-YN$_2$ monolayer with SOC calculated at the PBE and HSE06 levels. The blue lines indicate the PBE calculations, and the red lines indicate the HSE06 calculations.
(c) The enlarged band structures shown in the green rectangle in (b). The band gaps are indicated below the figure: $\Delta$E$_{PBE}$ = 29.7 meV and $\Delta$E$_{HSE06}$ = 97.5 meV.
}
\end{figure}

The space group of YN$_2$ is $P\overline{3}m1$ (No.164, D$_{3d}^3$), and its Wyckoff Positions are Y (0, 0, 0) and N (1/3, 2/3, $z$) with $z = 0.04489$.
As shown in Fig.~\ref{bz_bands}(a), the Y atom is the inversion center in the octahedral structure unit and the three N atoms in the upper layer will turn into the three N atoms in the lower layer under inversion.
The optimised lattice parameter and Y-N bond distance are 3.776 and 2.350 \AA, respectively, which is in good agreement with previous calculations \cite{Liu2017Nanores}.
Without SOC, we observe a distorted Dirac cone along the M-K high-symmetry line both at the PBE and HSE06 level.
Similar band structures can also be observed in other 2D non-magnetic systems, such as the TaCX \cite{Zhou2D2016} and the distorted hexagonal frameworks GaBi-X$_2$ (X = I, Br, Cl) \cite{Li2017nanores}. However, distorted Dirac cones in 2D magnetic systems are rare.
The calculated band structures with SOC are shown in Fig.~\ref{bz_bands}(b).
The blue lines indicate the band structure at the PBE level and a band gap (29.7 meV) can be observed at the Fermi level along the M-K high-symmetry line.
The semilocal approximations to the exchange-correlation energy at the PBE level underestimate the band gap with respect to experiment and overestimate electron delocalization effects for many $d$-element compounds.
Therefore, calculations at the HSE06 level are necessary to give more reliable band gaps and band structures.
We found that the band gap at the HSE06 level is larger than that at the PBE level.
To see the difference more clearly, the enlarged band structure are shown in Fig.~\ref{bz_bands}(c).
The calculated band gap at the PBE level is $\Delta$E$_{PBE}$ = 29.7 meV, while that at the HSE06 level $\Delta$E$_{HSE06}$ increases to 97.5 meV.
By projecting the wavefunctions onto the spherical harmonics, we analysed the occupation of the different atomic orbitals both with and without SOC near the Fermi level. Though there are minor contributions of $d$ orbitals from the Y atom near the Fermi level, the major contribution of the atomic orbitals comes from the $p$ orbitals of the N atoms. This is consistent with previous study \cite{Liu2017Nanores}.
Next, we constructed a tight-binding Hamiltonian with 12 Wannier functions by projecting the $p_x$, $p_y$, and $p_z$ orbitals of the two N atoms in the unit cell of 1T-YN$_2$ in order to further examine the electronic band structure.
Despite the difference in the band gap, the main character close to the Fermi level at the PBE and HSE06 level with SOC are qualitatively the same. In the following analysis, we will consider the results at the PBE level with SOC.

\begin{figure}[htb!]
\includegraphics[scale=0.3]{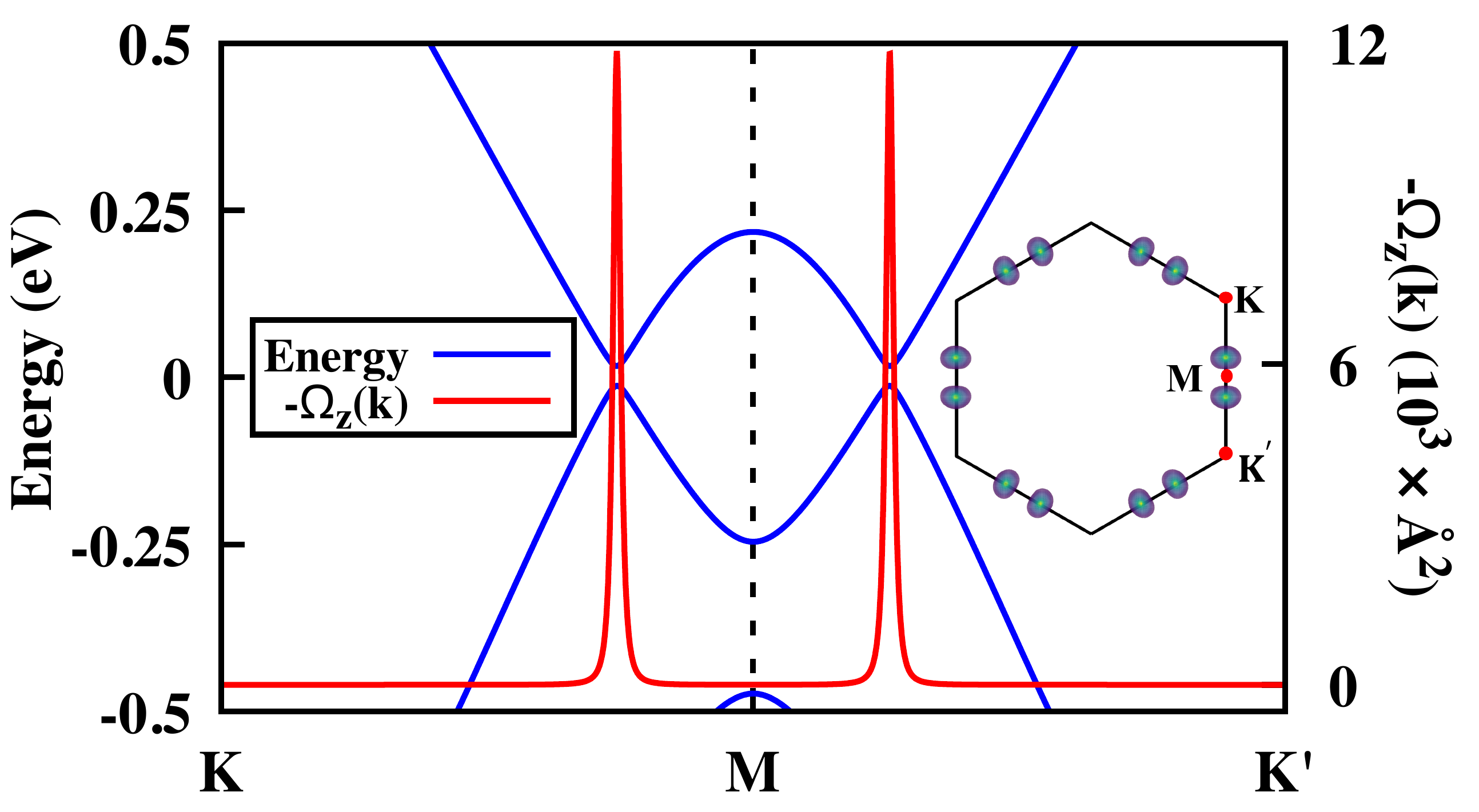}
\caption{
\label{bands_curv}
The reproduced band structures along the path K-M-K${'}$ (blue curves) with MLWF basis and the distribution of the corresponding Berry curvature in momentum space (red curves).
The inset indicates the position of the peaks of the Berry curvature in the 2D Brillouin zone.
}
\end{figure}

To investigate the topological properties of 1T-YN$_2$, we first calculated the gauge-invariant Berry curvature in momentum space.
The Berry curvature $\Omega_z(k)$ in 2D can be obtained by analyzing the Bloch wave functions from the self-consistent potentials:
\begin{eqnarray}
\Omega_z(k) = \sum_n f_n \Omega_n^z(k),
\end{eqnarray}
\begin{eqnarray}
\Omega_n^z(k)= -2\sum_{m \neq n} \text{Im} \frac{\langle \psi_n(k) | \upsilon_x | \psi_m (k) \rangle \langle \psi_m(k) | \upsilon_y | \psi_n (k) \rangle}{(\epsilon_m(k) - \epsilon_n(k))^2},
\end{eqnarray}
where $f_n$ is the Fermi-Dirac distribution function, $\upsilon_{x(y)}$ is the velocity operator, $\psi_n(k)$ is the Bloch wave function, $\epsilon_n$ is the eigenvalue and the summation is over all $n$ occupied bands below the Fermi level ($m$ indicates the unoccupied bands above the Fermi level).
In Fig.~\ref{bands_curv} the reproduced band structure (blue curves) and Berry curvature (red curves) along the high symmetry direction K-M-K${'}$, calculated by Wannier interpolation \cite{Mostofi20142309}, are shown.
To clarify that our MLWF is good enough to describe the electronic structures of 1T-YN$_2$, we also plot the band structures reproduced by the MLWF basis together with the PBE results along the high symmetry direction $\Gamma$-M-K in Fig. S1 in the supplementary information (SI).
As can be observed, the nonzero Berry curvature is mainly distributed around the avoided band crossings at the Fermi level.
The peaks in the Berry curvature at the two sides of the M point have the same sign. Furthermore, a plot of the Berry curvature peaks over the whole Brillioun zone (see inset of Fig.~\ref{bands_curv}) indicates that all 6 peaks have the same sign because of inversion symmetry.
Furthermore, the out of plane spin textures (see Fig. S2 in SI) is consistent with Berry curvature peaks, and we could conclude that the two Berry curvature peaks at the two sides of the M point contribute to the nonzero Chern number 1. The total Chern number from the 6 Berry curvature peaks is summed to 3, and this will be discussed below.

\begin{figure}[htb!]
\includegraphics[scale=0.55]{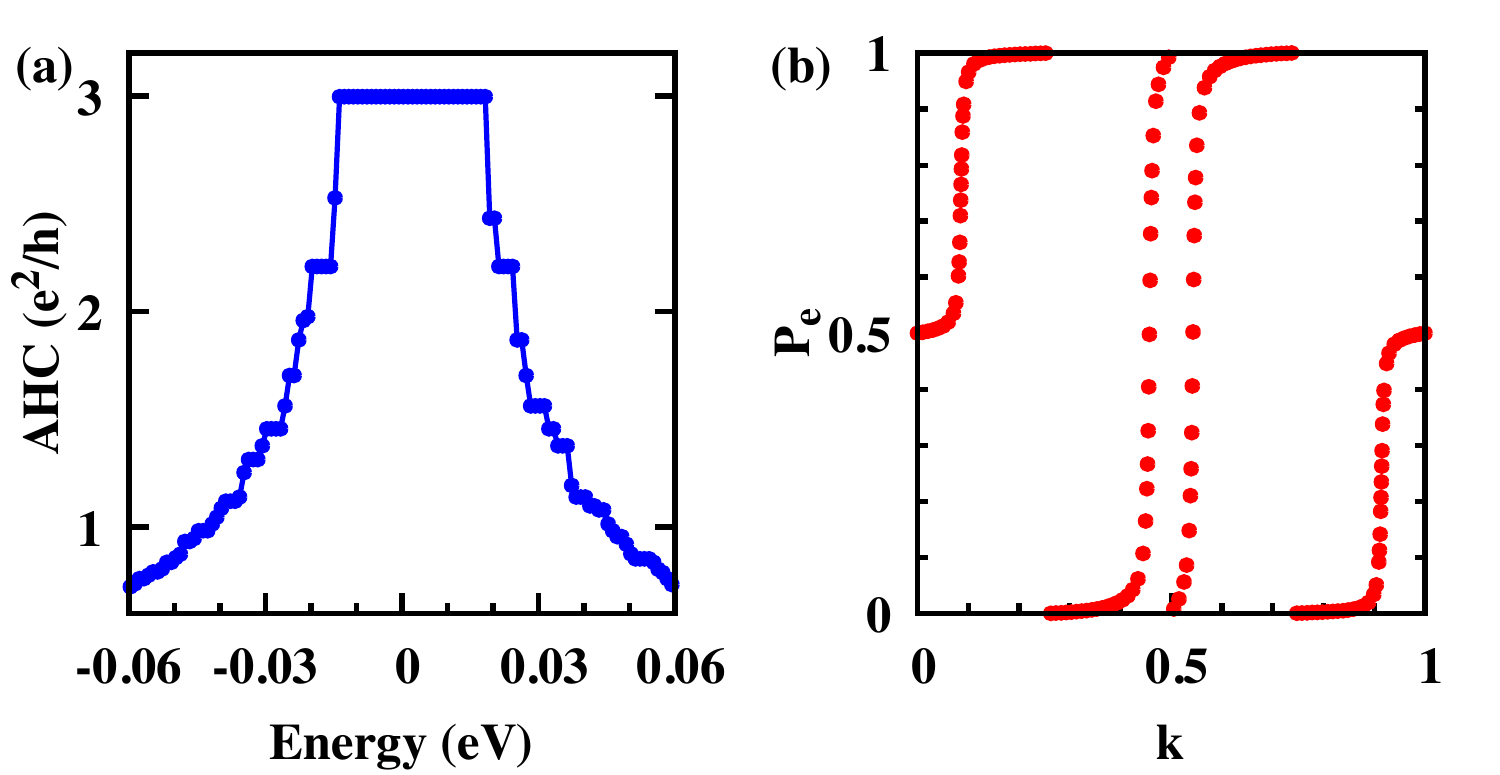}
\caption{
\label{ahc_wcc}
(a) The chemical-potential-resolved AHC when the Fermi level is shifted to zero.
(b) The change in the electronic polarization $P_e$.
}
\end{figure}

The Chern number is obtained by integrating the Berry curvature $\Omega_z(k)$ over the BZ,
\begin{eqnarray}
C = \frac{1}{2\pi} \int_{BZ}d^2k\Omega_z(k).
\end{eqnarray}
The Chern number $C$ is an integer and gives rise to the quantized Hall conductance: $\sigma_{xy} = Ce^2/h$.
$\sigma_{xy}$ is also known as the AHC, and the calculated chemical-potential-resolved AHC is shown in Fig.~\ref{ahc_wcc}(a).
A nontrivial gap of about 30 meV and a Chern number $C = 3$ can be deduced from the plateau near the Fermi level.
The non-zero Chern number can also be confirmed by evaluating the electronic polarization $P_e$ at discrete points in one primitive reciprocal lattice vector $k_y = k_i$ \cite{Vanderbilt2011PRBz2,YuR2011Prb,z2pack2017prb}.
In another primitive reciprocal lattice vector $k_x$, the hybrid Wannier charge centers (WCCs) can be defined as
\begin{eqnarray}
\overline{x}_n(k_y)=\frac{i}{2\pi} \int^{\pi}_{-\pi} dk_x \langle u_n(k_x,k_y) | \partial k_x | u_n(k_x,k_y)\rangle,
\end{eqnarray}
where $u_n(k_x,k_y)$ is the periodic part of the Bloch function $\psi_n (k)$.
The sum of the hybrid WCCs $\overline{x}_n$ will give the electronic polarization $P_e = e \sum_n \overline{x}_n (k_i)$ ($e$ stands for the electronic charge)  which is gauge invariant modulo a lattice vector.
This brings about a well-defined physical observable $\Delta P_e$ under a continuous deformation of the system.
Thus the Chern number is given by
\begin{eqnarray}
C = \frac{1}{e} \Delta P_e = \frac{1}{e}(P_e (2\pi) - P_e(0)).
\end{eqnarray}
As can be seen from Fig.~\ref{ahc_wcc}(b) which was calculated by WannierTools \cite{wanntools}, the electronic polarization $P_e$ shifts upwards with the winding number 3, so the Chern number $C = 3$.

\begin{figure}[htb!]
\includegraphics[scale=0.3]{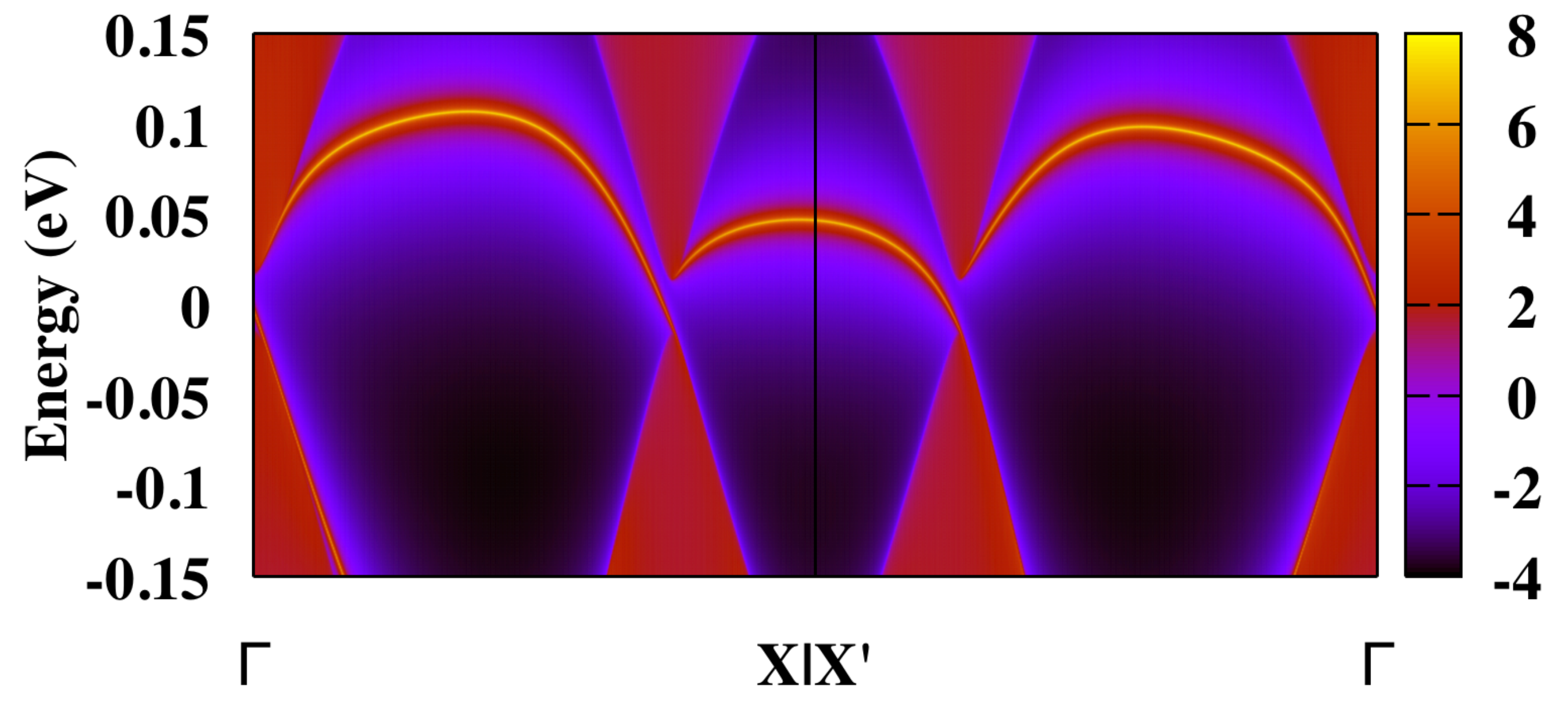}
\caption{
\label{sufdos}
Momentum and energy dependence of local density of states (log-scale) for the states at the edge of a semi-infinite plane ($\overline{1}$10).}
\end{figure}

According to the bulk-edge correspondence \cite{bulkedge1993PRL}, the non-zero Chern number is closely related to the number of nontrivial chiral edge states that emerge inside the bulk gap of a semi-infinite system.
With an effective concept of principle layers, an iterative procedure to calculate the Green's function for a semi-infinite system is employed.
The momentum and energy dependence of the local density of states at the edge can be obtained from the imaginary part of the surface Green's function:
\begin{eqnarray}
A(k,\omega)=-\frac{1}{\pi} \lim_{\eta \to 0^{+}} \text{Im} \text{Tr} G_s(k,\omega + i\eta),
\end{eqnarray}
and the results are shown in Fig.~\ref{sufdos}.
It is clear that there are three gapless chiral edge states that emerge inside the bulk gap connecting the valence and conduction bands and corresponding to a Chern number $C = 3$.

\section{Possibility of experimental realization}

\begin{figure}[htb!]
\includegraphics[scale=0.3]{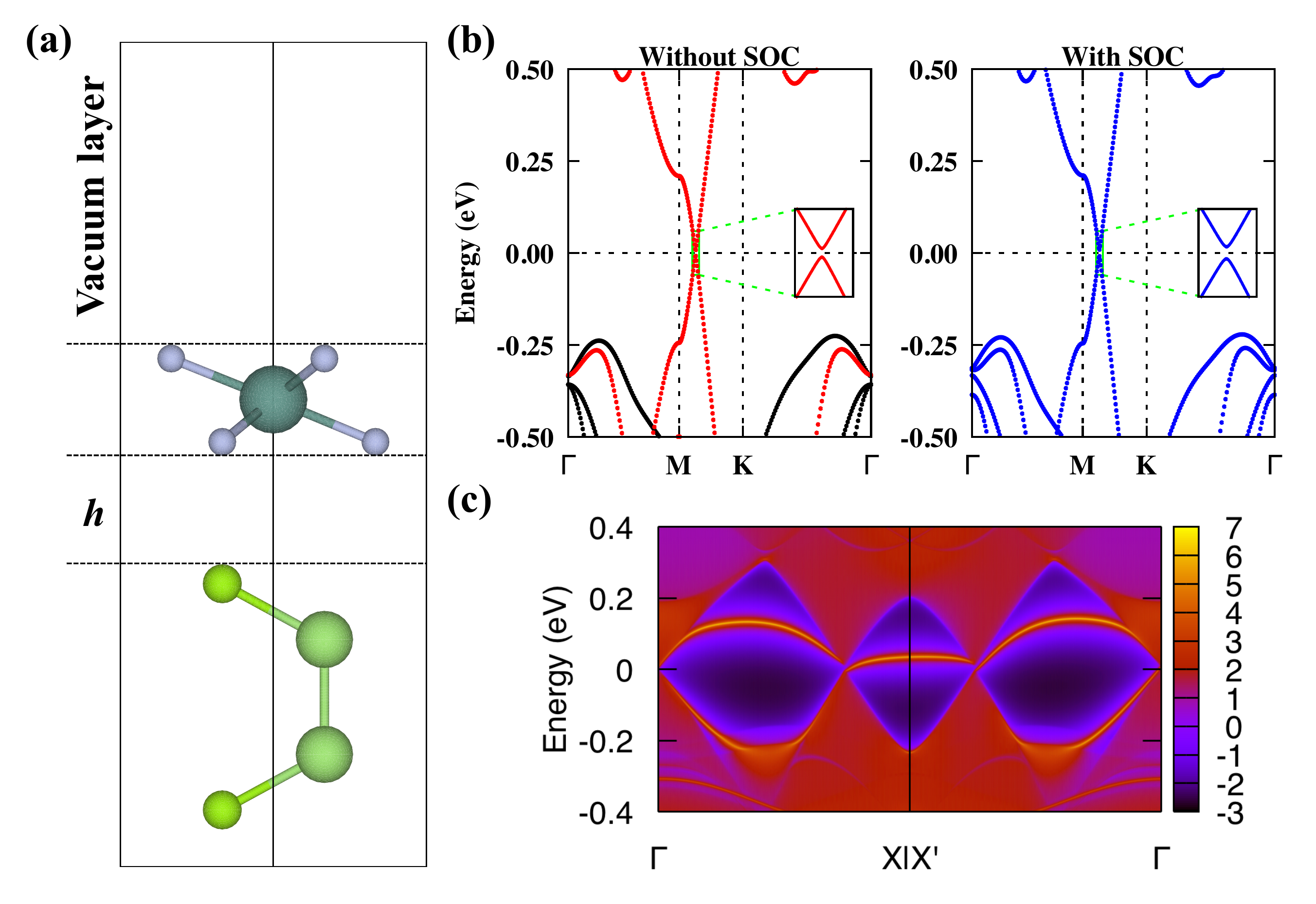}
\caption{
\label{heteroC}
(a) Side view of the YN$_2$/GaSe heterostructure with stacking model C along the (1$\overline{1}$0) direction. $h$ indicates the interlayer distance.
(b) The band structures of heterostructure C without SOC (left) and with SOC (right). The spin up band structure is plotted with black dots and the spin down band structure is given by red dots. The insets demonstrate the avoided band crossings.
(c) Momentum and energy dependence of local density of states for the states at the edge of heterostructure C.}
\end{figure}

In the following section, we will discuss the possibility of experimentally realizing the YN$_2$ monolayer. Notice that the lattice constant of 1T-YN$_2$ is close to that of the GaSe monolayer, which has been realized experimentally~\cite{acs2012gase} and therefore is expected to be a suitable substrate. We used the semiconducting GaSe monolayer as the substrate for the YN$_2$ monolayer. We found there are six kinds of highly symmetrical stacking models (A, B, C, D, E, and F) for the heterostructure (see Fig.~\ref{heteroC}(a) and also the red dashed rectangles in Figs. S3-S8 in SI), which are similar to those of the high-buckled silicene on MoS$_2$ substrate~\cite{JPCC2014li}.  First-principles calculations including vdW interactions resulted in a lattice constant of YN$_2$ of 3.765 \AA, which is slightly smaller than the PBE result (3.776 \AA) without vdW interactions. Comparing the lattice constant of the GaSe monolayer (3.804 \AA), the lattice mismatch between the two monolayers is only 1.0\%, which is far smaller than that of the high-buckled silicene on MoS$_2$ substrate (20.8\%)~\cite{Am2014silicene}, indicating its feasibility for experimental synthesis. Then we calculated the energy and the band structures of the six heterostructures without and with SOC effect. From Table~\ref{table}, we can clearly see that the energy is related to the interlayer distance ($h$) between the two monolayers. Stacking model B shows the lowest energy with the smallest distance $h$ and stacking model C/F shows the highest energy with the largest distance $h$. It can be easily understood that the smaller $h$, the stronger the interactions between the monolayers. All the six vdW YN$_2$/GaSe heterostructures show similar band structures as shown in Fig.~\ref{heteroC}(b) (also Figs. S3-S8). Not only the energy, but also the band gap without SOC is influenced by $h$. Comparing the other stacking models (A, B, D, and E), the band gap without SOC of the C/F is smaller, 16.6 meV, which indicates that the effect from the substrate is small. When considering SOC effect, changing the band gap with different size  is related with the stacking style. On the other hand, we found that stacking model C/F can keep its nontrivial topological properties with a Chern number $C = 3$, but other stacking models cannot keep the nontrivial properties ($C = 0$) due to the strong interaction with the substrate. To demonstrate the nontrivial topological properties, we calculated the corresponding edge states shown in Fig.~\ref{heteroC}(c) and Fig. S9. Notice that the stacking model C/F has similar three chiral edge states as the single 1T-YN$_2$ monolayer in the semi-infinite plane ($\overline{1}$10).

\begin{table*}
  \caption{
The DFT calculated total energy $E$, lattice constant $a$, interlayer distance $h$ between the YN$_2$ and GaSe layer, band gap $\Delta_{PBE}$ without SOC and $\Delta_{SOC}$ with SOC (at the Dirac point) of the six different stacked YN$_2$/GaSe heterostructures and GaSe/YN$_2$/GaSe quantum wells. All results are calculated at the PBE level. The first lines of every item correspond to the YN$_2$/GaSe heterostructures, and the second lines correspond to the GaSe/YN$_2$/GaSe quantum wells.
}
\label{table}
\begin{tabular}{c c c c c c c}
    \hline
    Stacking model  & A & B & C & D & E & F  \\
    \hline
    \hline
    \multirow{2}{*}{$E$ (eV)} & -37.450 & -37.470 & -37.302 & -37.353 & -37.357 & -37.300 \\
                    & -53.624 & -53.666 & -53.324 & -53.436 & -53.434 & -53.320\\
    \hline
    \multirow{2}{*}{$a$ (\AA)} & 3.800 & 3.800 & 3.776 & 3.781 & 3.783 & 3.776\\
                   &3.813 &3.813 &3.777 &3.789 &3.790 &3.780 \\
    \hline
    \multirow{2}{*}{$h$ (\AA)} & 2.402 & 2.338 & 3.030 & 2.739 &2.675 &3.037\\
                  &2.405 &2.341 &2.977 &2.748 &2.683 &3.028\\
    \hline
    \multirow{2}{*}{$\Delta_{PBE}$ (meV)} & 69.0 & 65.2 & 16.6 &62.8 & 69.0 & 16.6\\
                  & 0 & 0 & 0 & 0 & 0 & 0\\
    \hline
    \multirow{2}{*}{$\Delta_{SOC}$ (meV)} & 39.6 & 95.0 & 18.1 & 90.9 & 41.5 & 45.5\\
                   & 35.1 & 31.9 & 25.8 & 28.5 & 26.4 & 30.8\\ 
    \hline
\end{tabular}
\end{table*}

\begin{figure}[htb!]
\includegraphics[scale=0.3]{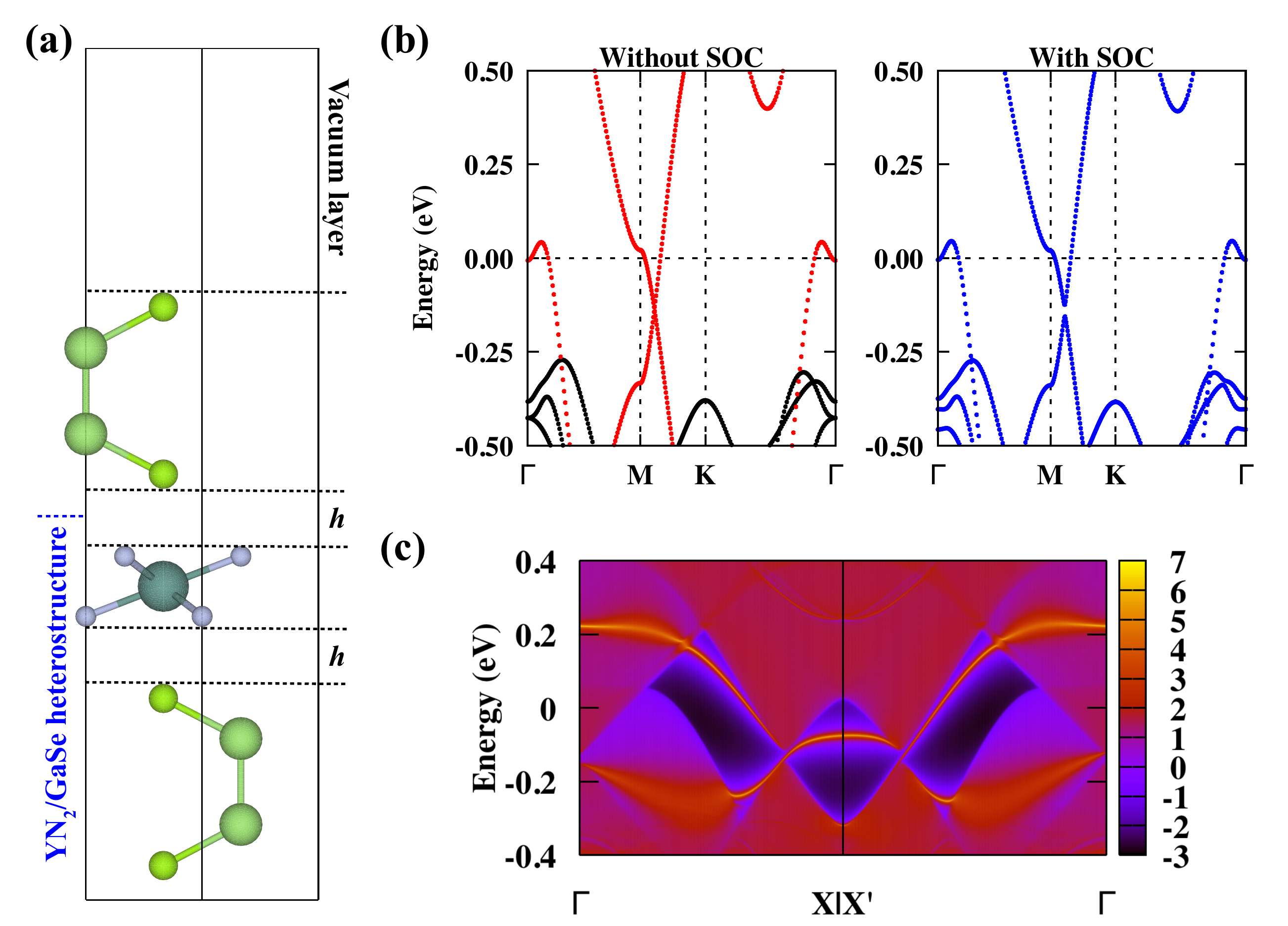}
\caption{
\label{wellB}
(a) Side view of the GaSe/YN$_2$/GaSe quantum well with stacking model B from the (1$\overline{1}$0) direction. $h$ indicates the interlayer distance.
(b) The band structures of quantum well B without SOC (left) and with SOC (right). The spin up band structure is plotted with black dots and the spin down structure is given by red dots.
(c) Momentum and energy dependence of local density of states for the states at the edge of quantum well B.}
\end{figure}

The nontrivial topological properties of the YN$_2$/GaSe heterostructure C/F is related with the interlayer distance ($h$), which may need some manipulations in experiment, and their energy is slightly higher than the other heterostructures. Here, we propose another way to keep the nontrivial properties of the YN$_2$. We construct GaSe/YN$_2$/GaSe quantum well structures by inverting the original GaSe monolayer in the YN$_2$/GaSe heterostructure about the Y atom in the YN$_2$ monolayer. Then, we obtain the sandwiched quantum wells as shown in Fig.~\ref{wellB} (see also Figs. S3-S8). The lowest energy quantum well structure is the stacking model B, which is shown in Fig.~\ref{wellB}(a). The corresponding band structures without and with SOC are shown in Fig.~\ref{wellB}(b). In the quantum well structure, the band gap influenced by the GaSe substrate is close to 0 (thus it retains the Dirac cone), and SOC can induce a nontrivial band gap (31.9 meV) in the Dirac cone, but the Dirac cone will submerge in the band line at the $\Gamma$ point, making the quantum well become semimetal. Due to the metallic bulk band structure, parts of the two edge states are absorbed by the bulk states as shown in Fig.~\ref{wellB}(c). Only one edge state through the X|X${'}$ point connecting the valence and conduction bands can be clearly seen. We could conclude that the GaSe/YN$_2$/GaSe quantum well structures maintain the nontrivial topological properties of YN$_2$ monolayer , and thus has the potential to be observed in experiments.

\begin{figure}[htb!]
\includegraphics[scale=0.4]{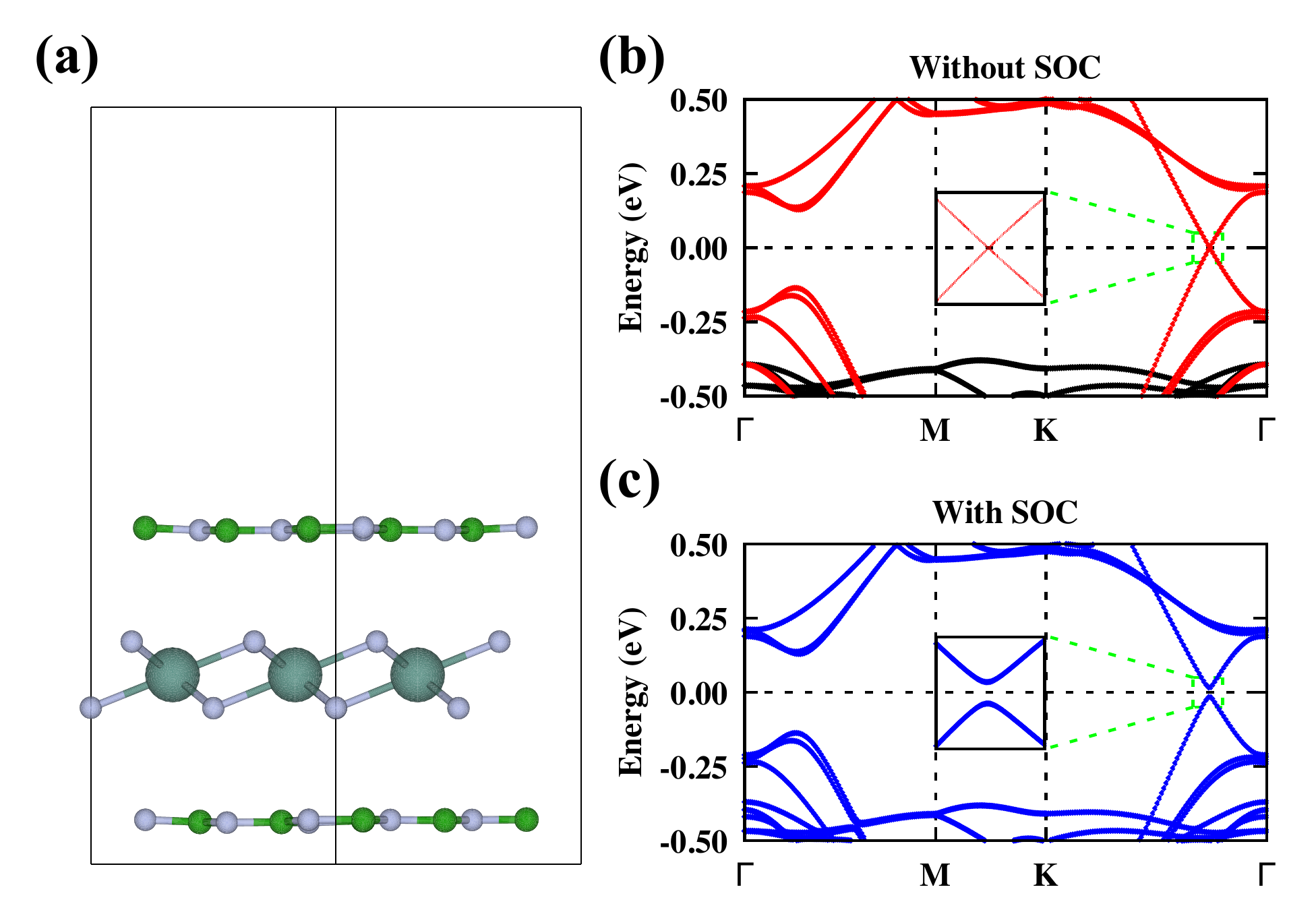}
\caption{
\label{hBN}
(a) Side view of hBN/YN$_2$/hBN quantum well along the (1$\overline{1}$0) direction.
(b) The band structures of  hBN/YN$_2$/hBN quantum well without SOC. The spin up band structure is plotted with black dots and the spin down band structure is given by red dots. The inset demonstrates the unavoided band crossing.
(c) The band structures of  hBN/YN$_2$/hBN quantum well with SOC. The inset demonstrates the avoided band crossing.
}
\end{figure}

Considering the Dirac cone preserved in the GaSe/YN$_2$/GaSe quantum well structure, we also propose hexagonal boron nitride (hBN) monolayer as a possible substrate to construct the hBN/YN$_2$/hBN quantum well structure (Fig.~\ref{hBN}(a) and Fig. S10). The hBN substrate has been a popular substrate for graphene in experiment~\cite{Dean2010} and silicene/germanene in theories~\cite{PRB2016hBN,LI20132628,pccp2013Li}, and the similar hBN/WTe$_2$/hBN quantum well has been realized experimentally for observing QSH effect at about 100 K\cite{wu2018observation}. The band structures of the hBN/YN$_2$/hBN quantum well structure are shown in Figs.~\ref{hBN}(b) and (c). Due to the band folding effect, the Dirac cone moves to the K-$\Gamma$ line as shown in Fig.~\ref{hBN}(b). Without SOC, the band gap is close to zero, and a band gap 32.6 meV is induced when considering the SOC effect. In contrast to the metallic GaSe/YN$_2$/GaSe quantum well, it is insulating with a sizable band gap. Thus, the hBN/YN$_2$/hBN quantum well is a good way for experimental synthesis of the YN$_2$ monolayer while keeping its topological nontrivial properties.

\section{Discussions}
By the calculations of Berry curvature, the AHC, the Chern number, and the corresponding edge states, we confirm the topological nontrivial properties of the 1T-YN$_2$ monolayer. 
Also, we construct YN$_2$/GaSe heterostructures, GaSe/YN$_2$/GaSe and hBN/YN$_2$/hBN quantum wells to investigate the effect of substrates, and we show that the nontrivial topological properties are preserved.
The most significant points of our research can be summarized as follows:
\begin{enumerate} [ (i) ]

\item The 1T-YN$_2$ monolayer has a nontrivial band gap of nearly 0.1 eV which is sufficient for the realization of QAH effect at room temperature. The found magnitude of the nontrivial band gap compares favorably to previous proposals \cite{Niu2016review,Yang2016cpb}. The largest band gap among the large number of QSH insulators is 1.08 eV \cite{Song2014NPG,Yaoprb2014}, but a band gap in QAH insulators of this order of magnitude has not been discovered. In recent reviews \cite{Niu2016review,Yang2016cpb}, some systems have been proposed exhibiting large nontrivial band gaps ($>$ 0.1 eV), but most of them are realized by functionalization methods, such as half functionalization of stanene with I atoms ($\approx$ 340 meV) \cite{YanBH2014PRL} and half functionalization of Bi (111) bilayer with H atoms ($\approx$ 200 meV) \cite{NiuCW2015PRB,LiuCC2015PRB}. However, their stability still needs to be confirmed, and realizing such functionalization seems to be difficult to control experimentally. In contrast, note that the stability of 1T-YN$_2$ has been demonstrated by energetic and dynamical analysis \cite{Liu2017Nanores}.
On the other hand, we demonstrate that GaSe and hBN monolayers are favorable substrates to construct heterostructures and quantum wells while keeping the nontrivial topological properties of 1T-YN$_2$ monolayer.
Since MoN$_2$ has been realized in experiment \cite{MON2015JACS} and its monolayer stability has been predicted \cite{MON2015Nanolett}, it seems likely that the realization of 1T-YN$_2$ monolayer is also possible.
\item Unlike the quantum Hall effect in which the Chern number can be tuned by changing the magnetic field or varying the Fermi level, the Chern number for intrinsic QAH effects in realistic materials are mostly limited to $\pm$ 1 and $\pm$ 2 \cite{Niu2016review,Yang2016cpb,SXL2014PRB}.
QAH insulators with high Chern number and corresponding nontrivial chiral edge states are rarely reported. These nontrivial edge states will provide strong currents and signals which are significant in experiments.
For example, more evidences should be observed in the QAH insulator/$s$-wave superconductor heterostructure for the chiral Majorana edge modes, which has been predicted to be abble to realize robust topological quantum computing~\cite{HeQL2017Science}.
Thus, the stable 1T-YN$_2$ monolayer with a high Chern number is expected to lead to important applications in experiments.
\item To our knowledge, this is the first time that the QAH effect is predicted in a 2D TMN material. At present, the topological QSH effect was predicted in 1T${'}$ TMDs \cite{Qian2014QSHE}. On the other hand, some TMD monolayers with square lattice \cite{MaYD2015PRB} and hexagonal lattice \cite{MaYD2016PRB} have also been predicted to be 2D topological insulators, but such structures have yet not been observed experimentally. Here, we report that the QAH effect can be realized in the 1T structure of TMN, which is a common structure for TMDs in exeperiment.
\end{enumerate}

\section{Conclusions}
In conclusion, we investigated the electronic band structure of 1T-YN$_2$ monolayer by first-principles calculations in the case of SOC and observed an intrinsic QAH effect. A large nontrivial band gap ( $\approx$ 0.1 eV) and high Chern number ($C = 3$) were found. We calculated the Berry curvature and AHC to demonstrate the nontrivial topological properties. Three nontrivial gapless chiral edge states were found which provide strong evidence for the realization of the QAH effect in experiments. 
Moreover, we propose different substrates, such as GaSe and hBN monolayers, for the possible experimental realization of the nontrivial topological properties in the  YN$_2$/GaSe heterostructures, GaSe/YN$_2$/GaSe and hBN/YN$_2$/hBN quantum wells.
The prediction of the QAH effect in the 1T-YN$_2$ monolayer provides a different type of structure and material for the investigation of  QAH insulators in the TM compounds.


\section*{Conflicts of interest}
There are no conflicts to declare.

\begin{acknowledgments}
This work is supported by Ministry of Science and Technology of China (MOST) (Grant No. 2016YFA0301604), National Natural Science Foundation of China (NSFC) (No. 11574008), the Thousand-Young-Talent Program of China, the Fonds voor Wetenschappelijk Onderzoek (FWO-Vl) and the FLAG-ERA project TRANS2DTMD. The computational resources and services used in this work were provided by the VSC (Flemish Supercomputer Center), funded by the Research Foundation - Flanders (FWO) and the Flemish Government - department EWI, and the National Supercomputing Center in Tianjin, funded by the Collaborative Innovation Center of Quantum Matter. W. Wang acknowledges financial support from the National Natural Science Foundation of China (Grant No. 11404214) and the China Scholarship Council (CSC).
\end{acknowledgments}

\bibliography{reference.bib}

\end{document}